\documentclass[8.5pt,twoside,twocolumn]{article}
\usepackage[super,sort&compress,comma]{natbib}
\usepackage{times,mathptmx}
\usepackage{sectsty}
\usepackage{balance}

\usepackage{graphicx} 
\usepackage{lastpage}
\usepackage[format=plain,justification=raggedright,singlelinecheck=false,font=small,labelfont=bf,labelsep=space]{caption}
\usepackage{fancyhdr}
\usepackage{float}
\begin{document}

\thispagestyle{plain}
\fancypagestyle{plain}{
\renewcommand{\headrulewidth}{1pt}}
\renewcommand{\thefootnote}{\fnsymbol{footnote}}
\renewcommand\footnoterule{\vspace*{1pt}%
\hrule width 3.4in height 0.4pt \vspace*{5pt}}
\setcounter{secnumdepth}{5}

\makeatletter
\def\subsubsection{\@startsection{subsubsection}{3}{10pt}{-1.25ex plus -1ex minus -.1ex}{0ex plus 0ex}{\normalsize\bf}}
\def\paragraph{\@startsection{paragraph}{4}{10pt}{-1.25ex plus -1ex minus -.1ex}{0ex plus 0ex}{\normalsize\textit}}
\renewcommand\@biblabel[1]{#1}
\renewcommand\@makefntext[1]%
{\noindent\makebox[0pt][r]{\@thefnmark\,}#1}
\makeatother
\renewcommand{\figurename}{\small{Fig.}~}
\sectionfont{\large}
\subsectionfont{\normalsize}

\fancyfoot{}
\fancyfoot[RO]{\footnotesize{\sffamily{1--\pageref{LastPage} ~\textbar  \hspace{2pt}\thepage}}}
\fancyfoot[LE]{\footnotesize{\sffamily{\thepage~\textbar\hspace{3.45cm} 1--\pageref{LastPage}}}}
\fancyhead{}
\renewcommand{\headrulewidth}{1pt}
\renewcommand{\footrulewidth}{1pt}
\setlength{\arrayrulewidth}{1pt}
\setlength{\columnsep}{6.5mm}
\setlength\bibsep{1pt}

\twocolumn[
  \begin{@twocolumnfalse}
\noindent\LARGE{\textbf{Formation of ultracold metastable Rb$_2$ molecules in their $v''=\,$0 level by blue-detuned photoassociation}}
\vspace{0.6cm}

\noindent\large{\textbf{M. A. Bellos,\textit{$^{a}$} D. Rahmlow,\textit{$^{a}$} R. Carollo,\textit{$^{a}$} J. Banerjee,\textit{$^{a}$} O. Dulieu,\textit{$^{b}$} A. Gerdes, \textit{$^{c}$} E. E. Eyler,\textit{$^{a}$} P. L. Gould,\textit{$^{a}$} and W. C. Stwalley\textit{$^{a}$}}}\vspace{0.5cm}\\
\noindent\textit{\small{\textbf{Submitted to PCCP on May 02, 2011}}}
\vspace{0.6cm}

\noindent \normalsize{We report on the observation of blue-detuned photoassociation in Rb$_2$, in which vibrational levels are energetically above the corresponding excited atomic asymptote. $^{85}$Rb atoms in a MOT were photoassociated at short internuclear distances to levels of the $1\,^3\Pi_g$ state at a rate of approximately $5\times10^4$ molecules/s. We have observed most of the predicted vibrational levels for all four spin-orbit components $0_g^+$, $0_g^-$, $1_g$, and $2_g$, including levels of the $0_g^+$ outer well. These molecules decay to the metastable $a\,^3\Sigma_u^+$ state, some preferentially to the $v''=\,$0 level, as we have observed for photoassociation to the $v'$=8 level of the $1_g$ component.}
\vspace{0.5cm}
 \end{@twocolumnfalse}
  ]

\footnotetext{\dag~Electronic Supplementary Information (ESI) available: [$0_g^+$, $0_g^-$, $1_g$, and $2_g$ potential energy curves]}
\footnotetext{\textit{$^{a}$~Department of Physics, University of Connecticut, Storrs, Connecticut 06269-3046, USA}}
\footnotetext{\textit{$^{b}$~Laboratoire Aim\'e Cotton, CNRS, Universit\'{e} Paris-Sud, b\^at. 505, 91405 Orsay, France}}
\footnotetext{\textit{$^{c}$~Computing and Storage Services, Regional Computer Centre for Lower Saxony, Leibniz Universit\"{a}t Hannover, Germany}}

\section{Introduction}

Photoassociation (PA)  of ultracold atoms is a powerful spectroscopic technique to produce and study ultracold molecules \cite{stwalley99, jones06}. Most photoassociation experiments access vibrational levels that are red-detuned from atomic transitions. Blue-detuned PA, where vibrational levels are energetically above their corresponding atomic asymptote, was first proposed to probe quasibound states \cite{dulieu97} and form ultracold Rb$_2$ \cite{almazor01} and KRb molecules \cite{skenderovic02}. Blue-detuned PA was first observed in Cs$_2$ \cite{pichler06} and Rb$_2$ \cite{weise09}.

There are two cases where blue-detuned photoassociation could occur: (1) in a local minimum of a potential energy curve repulsive at long range (as in the case here with the $1\,^3\Pi_g$ state of Rb$_2$) or (2) in a well that contains a potential barrier and vibrational levels above the atomic asymptote (for instance the $2^1\Sigma_g^+$ state in Rb$_2$ \cite{park01}). Although there is no fundamental difference between blue-detuned and red-detuned photoassociation, blue-detuned photoassociation generally occurs at small internuclear distances where the Franck-Condon factors for photoassociation are smaller. Furthermore blue-detuned photoassociation rates may be reduced by optical shielding effects \cite{marcassa94} where colliding atoms are prevented from reaching small internuclear distances. However an estimate based on Ref. \cite{Sanchez-Villicana95} indicates that these effects should be much less than one percent for our experimental configuration.
\begin{figure}[h]
\includegraphics[scale=0.7]{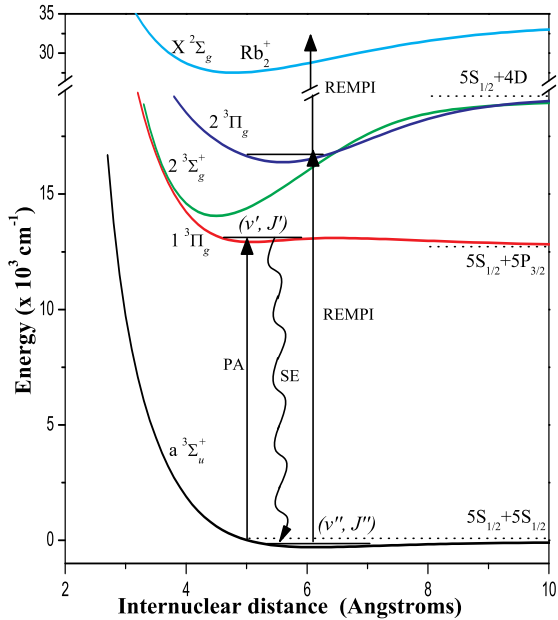}
\caption{\label{pec13pig} Scheme for producing and detecting ultracold metastable Rb$_2$ molecules. Blue-detuned photoassociation (PA) from free atoms to bound levels ($v', J'$) of the $1\,^3\Pi_g$ state is followed by spontaneous emission (SE) to a variety of ($v'', J''$) levels of the $a\,^3\Sigma_u^+$ state. Molecule detection through resonantly enhanced multiphoton ionization (REMPI) is a two step process achieved by first exciting molecules to an intermediate state ($2\,^3\Sigma_g^+$ or $2\,^3\Pi_g$), immediately followed by photoionization to produce Rb$_2^+$. The horizontal dotted lines indicate the positions of atomic asymptotes. The potential energy curves for Rb$_2$ and Rb$_2^+$ are from references \cite{lozeille06} and \cite{jraij03}, respectively.}
\end{figure}

Here, we demonstrate blue-detuned photoassociation to the $1\,^3\Pi_g$ state of $^{85}$Rb$_2$ as shown in Fig. \ref{pec13pig}. This is the first time a free-bound transition to the quasibound $1\,^3\Pi_g$ state has been directly observed. The $1\,^3\Pi_g$ state has been previously observed through transitions in a heat pipe oven \cite{almazor01,skenderovic02, veza98} and on liquid helium droplets \cite{mudrich09}. Photoassociation at short internuclear distance has been demonstrated \cite{deiglmayr09, deiglmayr10} on the B$\,^1\Pi$ state of LiCs, yielding ground rovibrational X$\,^1\Sigma^+$ state molecules \cite{deiglmayr08}.

\section{The $1\,^3\Pi_g$ potential energy curves}

\begin{figure}[h]
\includegraphics[scale=0.75]{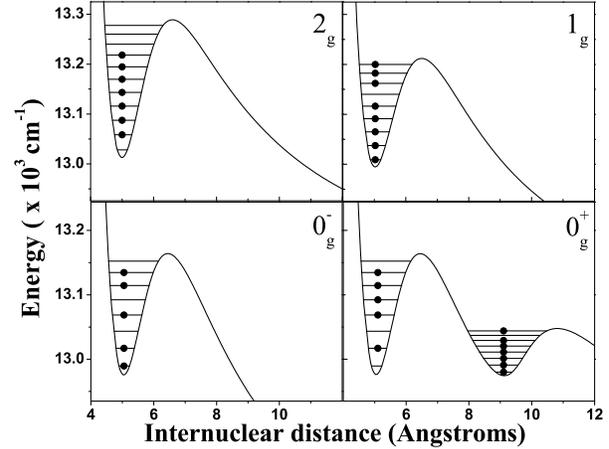}%
\caption{\label{foundstates} $1\,^3\Pi_g$ potential energy curves with spin-orbit coupling, along with vibrational levels calculated by the LEVEL 8.0 program\cite{level8}. Dots ($\bullet$) denote the experimentally observed vibrational levels. The levels that were unobserved most likely have a weaker photoassociation rate. The $0_g^+$ and $0_g^-$ state are double well systems, with an inner and outer well. The outer well of the $0_g^-$ state \cite{cline94,amiot95} (not shown) occurs at large internuclear distance and is red-detuned from the atomic asymptote.}
\end{figure}

Spin-orbit coupling splits the $1\,^3\Pi_g$ state into four distinct components, $1\,^3\Pi_{g,\,\Omega=0^+}$, $1\,^3\Pi_{g,\,\Omega=0^-}$, $1\,^3\Pi_{g,\,\Omega=1}$, and $1\,^3\Pi_{g,\,\Omega=2}$. $\Omega$ is the total electronic angular momentum projection on the internuclear axis, $g$ is the parity of the wavefunction by reflection through the center of mass, and (+/-) is the symmetry of the wavefunction by reflection through a plane containing the internuclear axis. These states can be expressed more compactly as $0_g^+$, $0_g^-$, $1_g$, and $2_g$ using Hund's case ({\it c}) notation. Their potential energy curves and bound levels are plotted in Fig. \ref{foundstates}. These potential curves are calculated using a rotation-based diabatization method within a quasidegenerate perturbation theory \cite{cimiraglia85}. In brief, the sixteen lowest adiabatic states of each relevant symmetry $^3\Pi_g$,  $^1\Pi_g$, $^3\Sigma_g^+$, $^1\Sigma_g^+$ in Hund's case ({\it a}) obtained by the method described in Ref.\cite{aymar05} are used as reference states at the internuclear distance of 40~a.u. (1~a.u.~=~0.527177~\AA), which are considered as representative of the separated-atom states with a reasonable accuracy. At this distance the potential energies including spin-orbit interaction are obtained after diagonalizing the Hamiltonian $H_{so}^{adia}=H^{adia}+H_{so}$, where the diagonal $H^{adia}$ matrix contains the adiabatic energies for all four symmetries above, and the coupling matrix $H_{so}$  the relevant atomic spin-orbit coupling terms for the dissociation limits up to $5^2S+6^2P$.

At each internuclear distance $R$ between 5~a.u. and 40~a.u., a rotation $\cal{R}$ of the subspace generated by the sixteen lowest adiabatic states is defined in order to maximize their overlap with the reference states above. This defines an effective hamiltonian $H^{eff}$ in an atomic-like basis, in which we introduce the $H_{so}$ matrix elements to set up a Hamiltonian matrix $H^{eff}_{so}$. The diagonalization of $H^{eff}_{so}$ at each $R$ yields potential curves including $R$-dependent spin-orbit couplings, such as those shown in Fig. \ref{foundstates}. Moreover, the inverse rotation $\cal{R}$$^{-1}$ of $H^{eff}_{so}$ back to the initial adiabatic states results in a non-diagonal matrix $H^{diab}_{so}$, where diagonal elements are the initial adiabatic potential curves, and off-diagonal terms the $R$-dependent spin-orbit couplings between these states. As noted in Ref.\cite{cimiraglia85}, the efficiency of the model is mainly limited by the overlap of the adiabatic states at $R$ with the reference states, which decreases from unity (at $R=40$~a.u. in the present case) to about 70\% at $R=10$~a.u.. As demonstrated in section 4, these results represent a good basis for the interpretation of the experimental measurements.

\section{Experiment}
\begin{figure}[h]
        \includegraphics[scale=0.55]{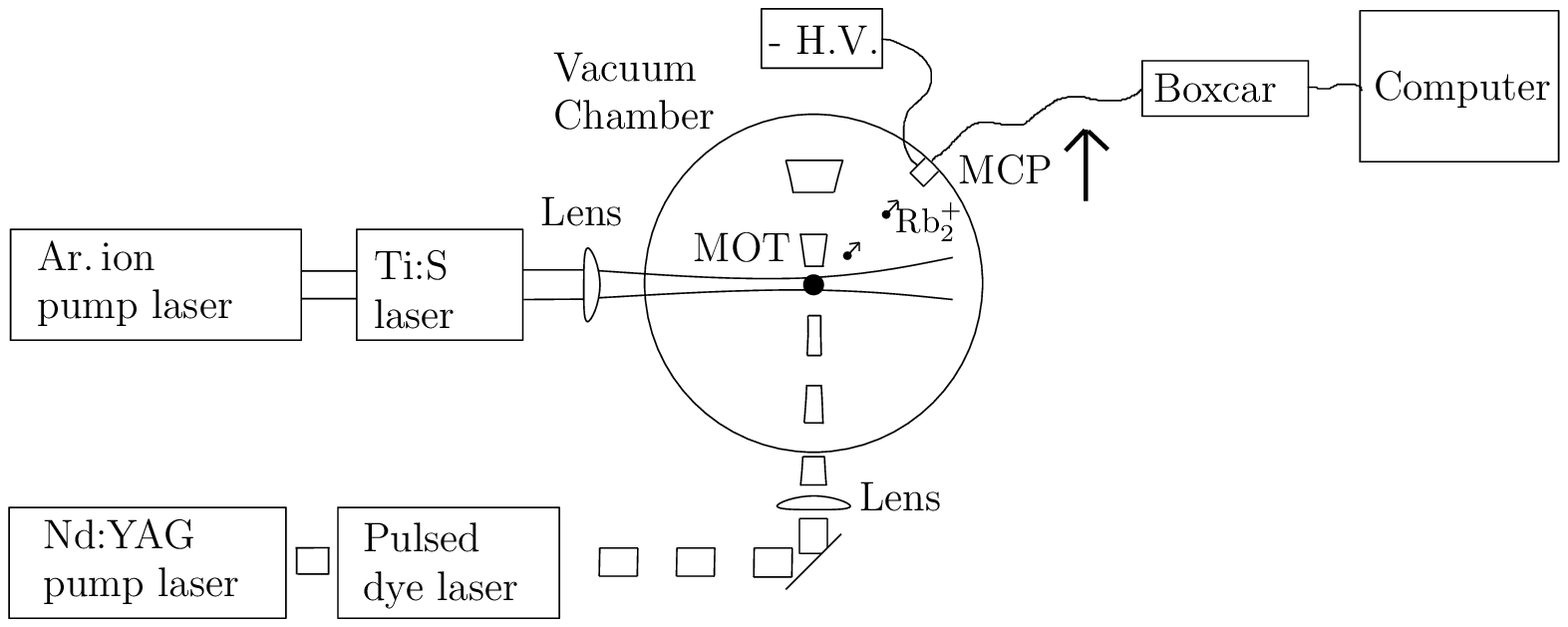} 
        \setlength{\unitlength}{0.02\linewidth}
        \begin{picture}(800,0)    
            \put(31,1){\includegraphics[width=20\unitlength]{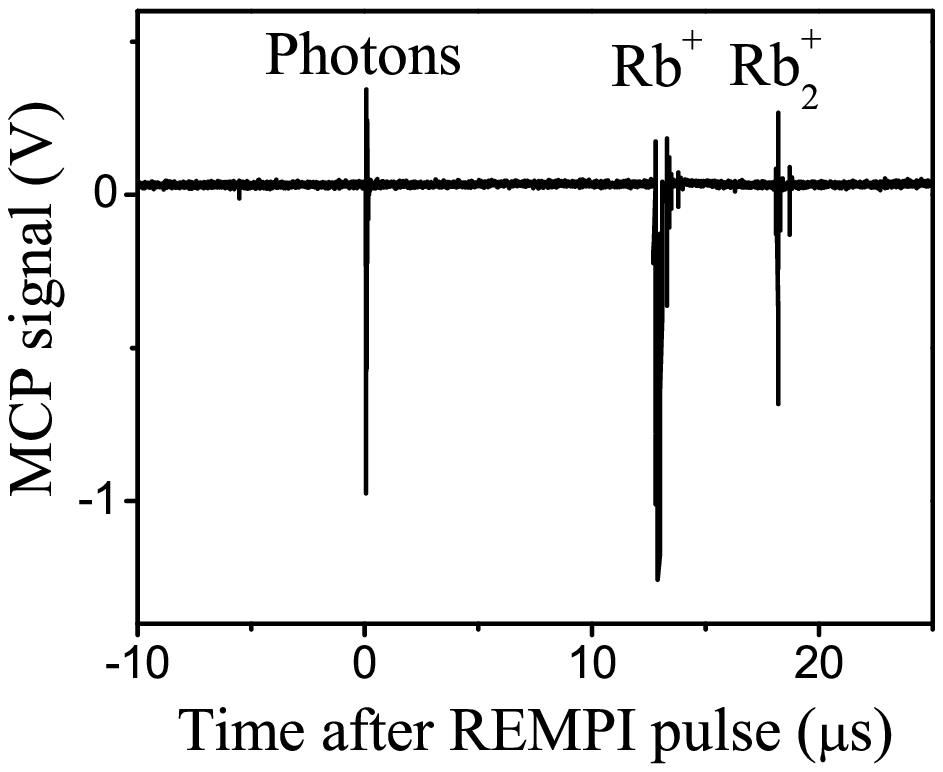}} 
        \end{picture}
        \caption{\label{setup} Schematic diagram of the photoassociation, REMPI and detection systems. The inset shows the time of flight of photons, atomic ions, and molecular ions reaching the MCP detector.}
    \end{figure}

The setup consists of an $^{85}$Rb magneto-optical trap (MOT) holding $\sim 10^6$ atoms at a temperature of $\sim 150\,\mu K$ with a density of $\sim\,10^{11}$ atoms/cm$^3$. The MOT trapping laser is locked 14 MHz below the $\mid5S_{1/2},F=3\rangle \rightarrow \mid5P_{3/2},F'=4\rangle$ transition at 780 nm. A repump laser locked on resonance to the $\mid5S_{1/2},F=2\rangle \rightarrow \mid5P_{3/2},F'=3\rangle $ transition is used to pump atoms that spontaneously decay to the $\mid5S_{1/2},F=2\rangle$ state back to the $\mid 5S_{1/2},F=3\rangle$ state. Since our optical repumping is not perfect, the energy splitting between the two hyperfine states $\mid5S_{1/2},F=2\rangle$ and $\mid5S_{1/2},F=3\rangle$ of 0.1012~cm$^{-1}$ is one that routinely appears in our PA spectra in the form of weak ``hyperfine ghost'' lines. We were not able to fully eliminate these atomic ``hyperfine ghost" lines from the spectra even after double checking for proper repump laser operation. A tunable cw Ti:sapphire laser (Coherent 899-29) with a power of $\sim\,$700~mW and linewidth of 500~kHz is focused approximately to the size of the MOT ($\sim\,$1~mm diameter) to photoassociate atoms into molecules as shown in Fig. \ref{setup}. The REMPI laser is a nanosecond pulsed dye laser (Continuum ND6000) with a pulse energy of $\sim\,$5$\,$mJ and linewidth of about 0.5 cm$^{-1}$ pumped by a Nd:YAG laser running at 532~nm with a 10~Hz repetition rate. The REMPI laser ionizes the atoms and molecules into Rb$^+$ and Rb$_2^+$, respectively. A boxcar averager integrates the ion signal within the time of flight range of Rb$_2^+$ ions. Our attempts to detect photoassociation by trap loss spectroscopy were unsuccessful; any decrease in MOT fluorescence was smaller than the fluorescence noise of the MOT.\\

\section{Photoassociation spectroscopy}
\begin{figure}[h]
\includegraphics[scale=0.7]{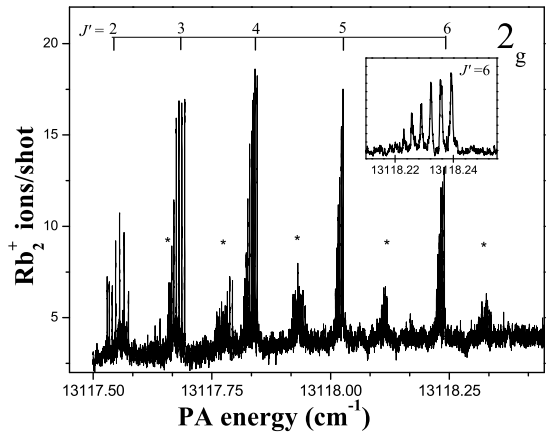}
\caption{\label{spectra2g} PA spectrum of the $2_g$ $v'$=2 level.  The inset shows a closeup of the $J'$=6 line showing molecular hyperfine structure. Rotational assignments are shown above the spectrum. Lines marked by an (*) are atomic ``hyperfine ghost" lines occurring 0.1$\,$cm$^{-1}$ above each rotational line.}
\end{figure}

\begin{figure}[h]
\includegraphics[scale=0.7]{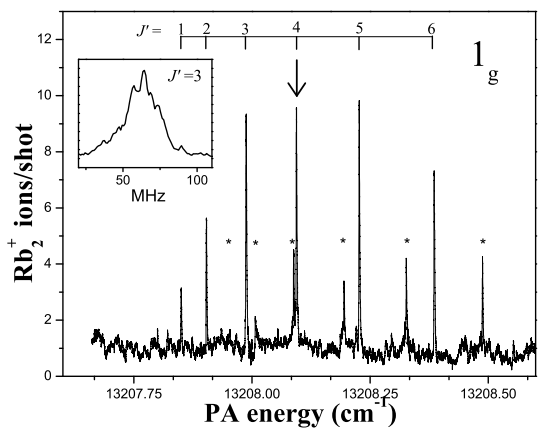}
\caption{\label{spectra1g} PA spectrum of the $1_g$ $v'$=8 level. Rotational assignments are shown above the spectrum. The arrow ($\downarrow$) indicates where the PA laser is fixed for the subsequent $a\,^3\Sigma_u^+$ $v''$=0 REMPI spectrum. The inset shows a closeup of the $J'=$3 line where the molecular hyperfine lines are mostly unresolved. Lines marked by an (*) are atomic ``hyperfine ghost" lines occurring 0.1$\,$cm$^{-1}$ above each rotational line.}
\end{figure}

\begin{figure}[h]
\includegraphics[scale=0.7]{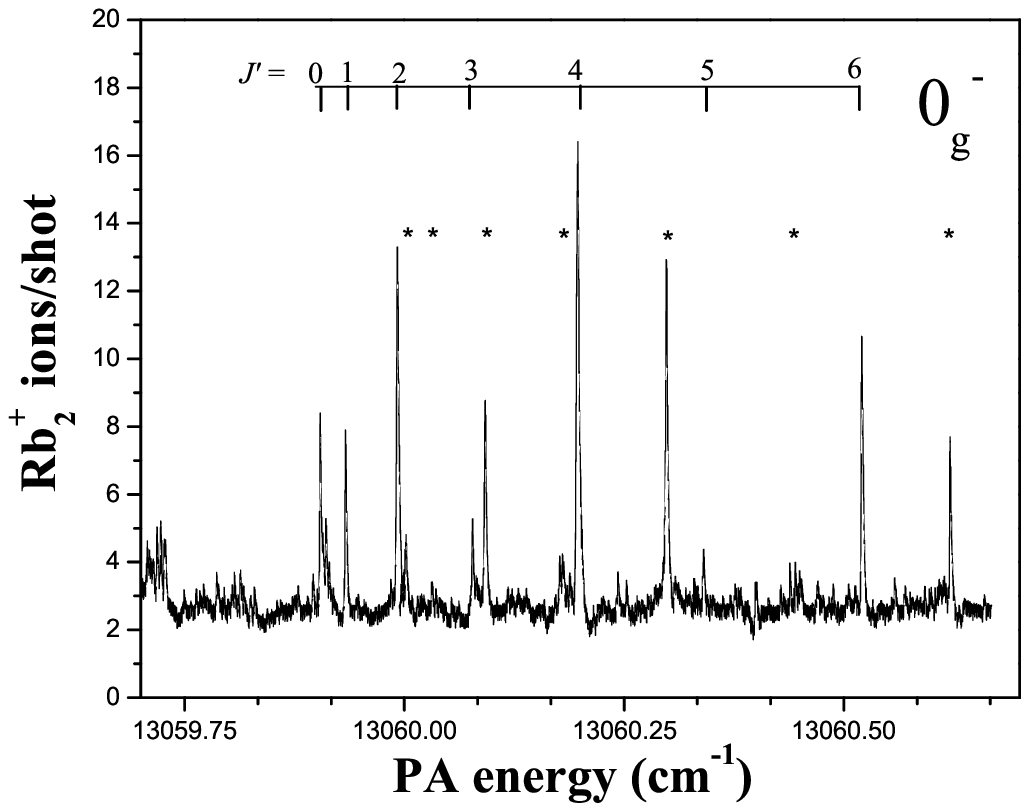}
\caption{\label{spectra0g-} PA spectra of the $0_g^-$ $v'$=3 level. Rotational assignments are shown above the spectra. Lines marked by an (*) are atomic ``hyperfine ghost" lines occurring 0.1$\,$cm$^{-1}$ above each rotational line.}
\end{figure}

\begin{figure}[h]
\includegraphics[scale=0.7]{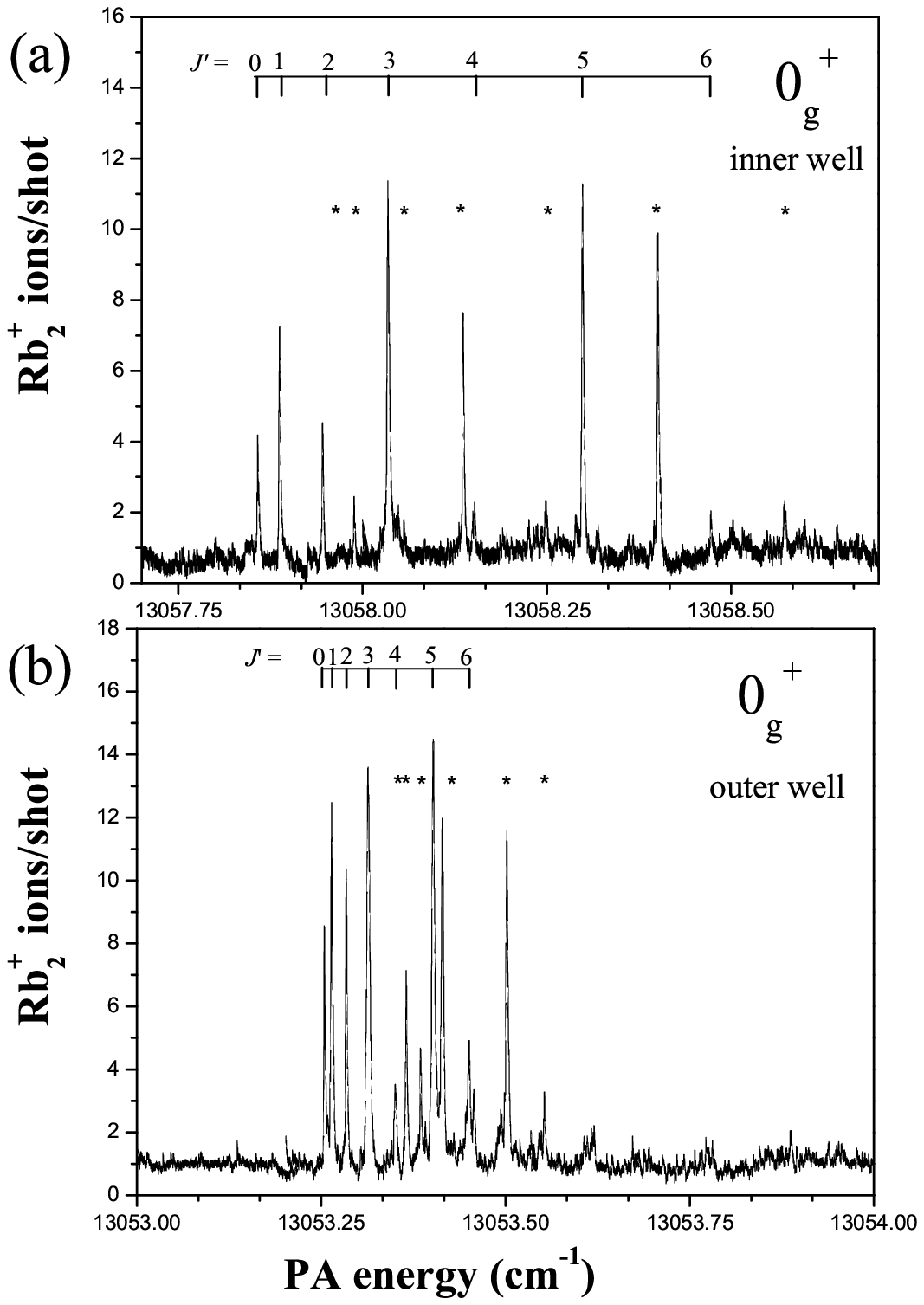}
\caption{\label{spectra0g+} PA spectra of the $0_g^+$ inner well $v'$=3 level (a) and the $0_g^+$ outer well $v'$=5 level (b). Rotational assignments are shown above the spectra. Lines marked by an (*) are atomic ``hyperfine ghost" lines occurring 0.1$\,$cm$^{-1}$ above each rotational line.}
\end{figure}

PA spectra were obtained by scanning the PA laser while monitoring the production of Rb$_2^+$ formed by the REMPI laser. Typical photoassociation spectra are shown in Figures \ref{spectra2g}, \ref{spectra1g}, \ref{spectra0g-}, and \ref{spectra0g+}. Each spectrum shows rotational lines, atomic ``hyperfine ghost" lines, and in some cases molecular hyperfine lines.

If $\Omega>0$, the electronic angular momentum can couple with the nuclear angular momentum, resulting in molecular hyperfine splittings. Hyperfine splittings are therefore expected for the $2_g$ and $1_g$ states, but not for the $0_g^+$ and $0_g^-$ states. We were able to resolve the molecular hyperfine splitting for $2_g$ states (inset of Fig. \ref{spectra2g}), but not for $1_g$ states (inset of Fig. \ref{spectra1g}).

The $0_g^-$ state (Fig. \ref{spectra0g-}) has stronger lines for even rotational quantum numbers, while the $0_g^+$ state in both inner and outer wells (Fig. \ref{spectra0g+}) has stronger lines for odd rotational quantum numbers. The $0_g^+$ outer well levels are identified by their smaller rotational constants $B_v$.

Our data shows rotational lines up to a maximum of $J'$=6. The data also shows that $J'>$3 lines are generally stronger than $J'<$3 lines. For red-detuned photoassociation using the same experimental set-up, we also observe rotational lines up to J'=6, however the lines with $J'>$3 are much weaker than lines with $J'<$3. The presence of blue-detuned light may have a heating effect on the MOT, which would favor transitions to higher rotational levels. We are currently investigating this possibility.


The energy of a rovibrational level is given to first order by $E_{v, J}= T_v+B_v[J(J+1) - \Omega^2]$. Here $T_v$ and $B_v$ are the term energy and the rotational constant of a vibrational level, respectively. $J$ is the rotational quantum number, which is always greater than or equal to $\Omega$. We extract the experimental rotational constant $B_v^{EXP}$ by fitting a straight line to the energy of the rotational lines versus $J(J+1) - \Omega^2$, the results of which are tabulated in Table \ref{comparison}. This fitting process also allowed us to assign an $\Omega$ quantum number to each spectrum. The experimental term energies $T_v^{EXP}$ are simply the wavenumbers of the photoassociation laser plus the average thermal energy of collisions ($\sim 10^{-4}$ cm$^{-1}$). The theoretical rotational constant $B_v^{THE}$ and term energy $T_v^{THE}$ are derived from the \textit{ab-initio} potential energy curves using the LEVEL 8.0 program\cite{level8}. Assigning the vibrational quantum numbers to the spectra was greatly simplified by knowledge of the theoretical vibrational energy spacings. After assigning the vibrational numbers, we were able to determine the energy shift to the potential curves necessary make them match the experiment. These energy shifts were -61, -27, -61, -51, and -100~cm$^{-1}$ for the $0_g^+$, $0_g^+$ outer well, $0_g^-$, $1_g$, and $2_g$ states, respectively. The unshifted potential energy curves can be found as electronic supplementary information\dag.

The areas under specific rovibrational lines of the $1\,^3\Pi_g$ state are also listed in Table \ref{comparison}. This area is proportional to the product of the photoassociation rate and the ionization rate, the latter of which depends on the frequency of the REMPI laser. After varying the REMPI frequency for many of the PA scans in an effort to obtain the strongest Rb$_2^+$ signal, we expect the reported line areas to roughly approximate the photoassociation rate. If one compares the calculated\cite{almazor01} photoassociation rate to the $0_g^+$ outer well with these measured line areas, one can see similarities; namely, an increase in PA rate with vibrational level followed by strong oscillations.

\begin{table}[h]
\scriptsize
  \caption{Experimental and theoretical rotational constants ($B_v$) and vibrational term energies for $J'$=3 ($T_{v,\,J'=3}$) for levels of the $1\,^3\Pi_g$ state in units of cm$^{-1}$. The theoretical term energies are shifted (see text) to match the experimental term energy of the lowest observed vibrational level. The area (A) under PA spectral lines for $J'$=3 of the $2_g$, $1_g$, and $0_g^+$ states, and $J'$=4 of the $0_g^-$ state in arbitrary units. This line area is an approximation of relative photoassociation rates.}
  \label{comparison}
  \begin{tabular*}{0.48\textwidth}{lllllll}
    \hline
   \small{State} &  \small{$v'$} &  \small{$B_{v}^{EXP}$} &  \small{$B_{v}^{THE}$} &  \small{$T_{v,\,J'=3}^{EXP}$}& \small{$T_{v,\,J'=3}^{THE}$}& \small{A} \\
    \hline
    $2_g$   &	0	&	-	        &	0.01583	&	-       	&	13029.293 &-	\\
            &	1	&	0.0152(9)	&	0.01568	&	13059.43(1)	&	13059.433 &1.5	\\
            &	2	&	0.01513(3)	&	0.01550	&	13089.04(1)	&	13088.595 &1.0	\\
            &	3	&	0.0156(3)	&	0.01530	&	13117.68(1)	&	13117.029 &14	\\
            &	4	&	0.0152(3)	&	0.01510	&	13145.44(1)	&	13144.457 &1.1	\\
            &	5	&	0.01470(1)	&	0.01488	&	13172.06(1)	&	13170.650 &13	\\
            &	6	&	0.0143(2)	&	0.01461	&	13197.47(1)	&	13195.609 &1.2	\\
            &	7	&	0.01382(8)	&	0.01431	&	13221.54(1)	&	13219.201 &15	\\
            &	8	&	-	        &	0.01395	&	-	        &	13241.166 &-	\\
            &	9	&	-	        &	0.01349	&	-       	&	13261.174 &-	\\
            &	10	&	-	        &	0.01278	&	-          	&	13278.579 &-	\\
                                                                                \\
    $1_g$   &	0	&	0.0158(2)	&	0.01561	&	13008.610(1)	&	13008.610 &8.6	\\
            &	1	&	0.0154(2)	&	0.01543	&	13037.791(1)	&	13037.044 &1.7	\\
            &	2	&	0.01533(3)	&	0.01523	&	13065.957(1)	&	13064.479 &14	\\
            &	3	&	0.01494(6)	&	0.01500	&	13093.040(1)	&	13090.810 &0.8	\\
            &	4	&	0.0147(2)	&	0.01475	&	13119.053(1)	&	13115.910 &7.2	\\
            &	5	&	-	&	0.01446	&	-	&	13139.627&-	\\
            &	6	&	0.01423(1)	&	0.01411	&	13166.936(1)	&	13161.752 &14	\\
            &	7	&	0.01370(2)	&	0.01366	&	13188.488(1)	&	13181.949 &0.7	\\
            &	8	&	0.01338(6)	&	0.01299	&	13207.987(1)	&	13199.580 &4.8	\\

                                                                                    \\
    $0_g^-$ &	0	&	0.0151(3)	&	0.015490	&	12980.840(1)	&	12980.840 &1.9	\\
            &	1	&	0.01507(3)	&	0.015288	&	13008.388(1)	&	13008.264 &8.0	\\
            &	2	&	-	        &	0.015062	&	-              	&	13034.674 &-	\\
            &	3	&	0.01465(5)	&	0.014816	&	13060.092(1)	&	13059.846 &7.5	\\
            &	4	&	-         	&	0.014534	&	-             	&	13083.558 &-	\\
            &	5	&	0.01408(2)	&	0.014187	&	13106.164(1)	&	13105.664 &9.4	\\
            &	6	&	0.01367(6)	&	0.013742	&	13126.527(1)	&	13125.874 &3.6	\\
            &	7	&	-         	&	0.013070	&	-             	&	13143.518 &-	\\
                                                                                        \\
$0_g^+$             &	0	&	-   &	0.015489	&	-             	&	12979.282 &-	\\
inner               &	1	&	0.01510(3)	&	0.015286	&	13006.693(1)	&	13006.693 &5.3	\\
well                &	2	&	-        	&	0.015058	&	-              	&	13033.079 &-	\\
                    &	3	&	0.01465(2)	&	0.014812	&	13058.035(1)	&	13058.223 &10	\\
                    &	4	&	0.0141(1)	&	0.014530	&	13081.793(1)	&	13081.910 &1.9	\\
                    &	5	&	0.01396(5)	&	0.014182	&	13104.167(1)	&	13103.983 &7.0	\\
                    &	6	&	0.01364(9)	&	0.013733	&	13124.408(1)	&	13124.144 &2.2	\\
                    &	7	&	-         	&	0.013049	&	-             	&	13141.709 &-	\\
                                                                                                \\
$0_g^+$             &	0	&	0.00478(8)	&	0.004791	&	13005.612(1)	&	13005.612  &0.8	\\
outer               &	1	&	0.00463(8)	&	0.004828	&	13016.113(1)	&	13016.556  &0.3	\\
well                &	2	&	0.00478(5)	&	0.004851	&	13026.170(1)	&	13026.942  &4.0	\\
                    &	3	&	0.00463(7)	&	0.004859	&	13035.708(1)	&	13036.788  &12	\\
                    &	4	&	0.00481(2)	&	0.004851	&	13044.764(1)	&	13046.068  &5.1	\\
                    &	5	&	0.00504(9)	&	0.004818	&	13053.313(1)   	&	13054.730   &18	\\
                    &	6	&	-         	&	0.004745	&	-             	&	13062.674    &-	\\
                    &	7	&	0.00472(2)	&	0.004561	&	13068.586(1)	&	13069.629  &6.5	\\
    \hline
  \end{tabular*}
\end{table}

\begin{table}[h]
\footnotesize
 \caption{Energy splitting between $(v'=1, J'=3)$ levels for various states. The theoretical splitting are derived from the potential curves given in the ESI. The experimental splitting is derived from Table \ref{comparison}}
 \label{comparison2}
  \begin{tabular*}{0.5\textwidth}{llll}
    \hline
    \small{Splitting} & \small{Splitting Type} & \small{Theory} & \small{Experiment} \\
                      &              & [cm$^{-1}$]    &  [cm$^{-1}$] \\
    \hline
       $2_g-1_g$ &Spin-orbit & 70.1 & 21.64(1) \\
       $1_g-0_g^-$ &Spin-orbit  & 19.2 &29.403(1)\\
       $0_g^+$$_{\textnormal{inner well}}-0_g^+$ $_{\textnormal{outer well}}$ &Well position  & 26.7 & -9.420(1)\\
        $0_g^--0^+_{g}$$_{\textnormal{inner well}}$ & Reflection symmetry & -0.5 & 1.695(1)\\
    \hline
  \end{tabular*}
\end{table}

The lifetimes of levels of the $1\,^3\Pi_g$ state can be reduced by tunneling through the potential barrier. This tunneling corresponds to a molecule dissociating into two free atoms, and is more likely to happen for higher vibrational levels where the barrier is less high and narrower. The tunneling lifetime has been calculated for the analogous state in KRb \cite{kim09} (the $2\,^3\Pi$ state) and varies between quasi-infinite lifetimes for the $v'$=0 level and $4\times 10^{-12}$~s for the uppermost vibrational level. For Rb$_2$, the calculated lifetime for the uppermost level of the $0_g^+$ outer well \cite{dulieu97} is 0.054ns. This lifetime correspond to a linewidth of 2950 MHz. So far we have not been able to observe strong broadening of the higher vibrational levels. In particular, we measure a total linewidth for the $v'$=7 level of the $0_g^+$ outer well of less than 50 MHz, and for the $v'$=8 level of the $1_g$ state, less than 25 MHz.

The quality of the present theoretical model for molecular spin-orbit can be assessed by looking at the energy shift of potential curves reported above, and the energy splitting between potential curves listed in Table \ref{comparison2}. In the distance range of the inner wells, the potential curves are very similar to the $1^3\Pi_g$ one, but split into different spin-orbit components. First of all, it is well known that this kind of quantum chemistry calculation usually predicts potential well depths like the one of the $1^3\Pi_g$ curve with an accuracy of about 100~cm$^{-1}$. The shifts reported above are thus consistent with this accuracy. The shifts are not the same for all curves, as they involve different Hund's case ({\it a}) curves with various individual accuracies: the $0_g^+$, $0_g^-$, $1_g$ curves result from the coupling between $^3\Pi_g$  and $^1\Sigma_g^+$, between $^3\Pi_g$ and $^3\Sigma_g^+$, between $^3\Pi_g$, $^1\Pi_g$ and $^3\Sigma_g^+$ respectively, while the $2_g$ curve involve only $^3\Pi_g$. Nevertheless, it is encouraging that the inner wells of all the states must be shifted by about the same amount to match the position of the experimental levels as described in Table \ref{comparison}. Due to the form of $H_{so}$ \footnote{The expression of the $H_{so}$ matrix for states correlated to an $^2S+^2P$ dissociation limit are displayed for instance in Ref.\cite{beuc07}, and those for the $^2S+^2D$ case in Ref.\cite{lozeille06}.}, the potential well of the $1_g$ curve is almost unshifted compared to the one of the original $1^3\Pi_g$ curve, as the shift of -51~cm$^{-1}$ (the smallest among inner wells) illustrates. The shift is different for the inner and outer wells of the $0_g^+$ curve, which is expected as the spin-orbit model is more accurate for large internuclear distances. The spin-orbit splitting between $1_g$ and $2_g$ curves (\textit{i.e.} the spin-orbit coupling diagonal matrix element in $H^{diab}_{so}$ for the $2_g$ symmetry) is overestimated by about 49~cm$^{-1}$. In contrast, that for the $0_g^+$ and $0_g^-$ symmetries is underestimated by 10~cm$^{-1}$. Finally, the tiny splitting between the $0_g^+$ and $0_g^-$ curves can only be predicted, at best, to the right order of magnitude. This is not surprising, as that energy splitting is much smaller than other energy splittings and beyond the accuracy of the present spin-orbit model.

\section{REMPI spectroscopy}
Molecules in the $1\,^3\Pi_g$ state spontaneously decay predominantly to the $a\,^3\Sigma_u^+$ state. There is little spontaneous decay to the ground electronic state $X\,^1\Sigma_g^+$ due to the electric dipole (E1) selection rules for spin ($\Delta S=0$) and parity ($u\rightarrow g$).

\begin{figure}[h]
\includegraphics[scale=0.75]{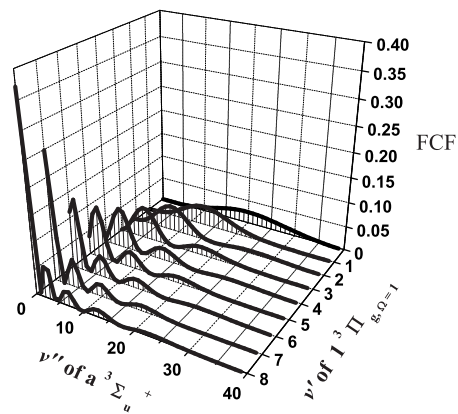}%
\caption{\label{fcf1g}  Franck-Condon factors for spontaneous emission between the $1\,^3\Pi_{g,\,\Omega=1}$ and $a\,^3\Sigma_u^+$ states. These calculations from LEVEL8.0 are based on the $1_g$ potential presented here and the $a\,^3\Sigma_u^+$ potential \cite{strauss10}.}
\end{figure}

The Franck-Condon factors (FCFs) for emission from $1_g$ to $a\,^3\Sigma_u^+$ are shown in Fig. \ref{fcf1g}; the largest FCF=0.37 is between the $v'$=8 and $v''$=0 levels. Therefore $\sim$37\% of the molecules in the $v'$=8 level should decay to the $v''$=0 level, $\sim$31\% to all other vibrational levels $v''$=1 to $v''$=39, and the remaining $\sim$32\% to bound-free transitions that produce free atoms. The FCF's for emission from $2_g$, $0_g^+$ inner well, and  $0_g^-$ to the $a\,^3\Sigma_u^+$ state all have a similar distribution to the one plotted in Fig. \ref{fcf1g}. The highest FCF from $2_g$ levels to $v''$=0 is 30\% starting from the $v'$=9 level. The highest FCF from the $0_g^+$ inner well and $0_g^-$ levels is 40\% starting from the $v'$=7 level. The $0_g^+$ outer well decays almost entirely between the $v''$=15 to $v''=$30 levels, regardless of the starting vibrational level $v'$. Population of the $a\,^3\Sigma_u^+$ $v''$=0 level has previously been achieved \cite{lang08} though the technique of magnetoassociation followed by STIRAP transfer.

\begin{figure}[h]
\includegraphics[scale=0.75]{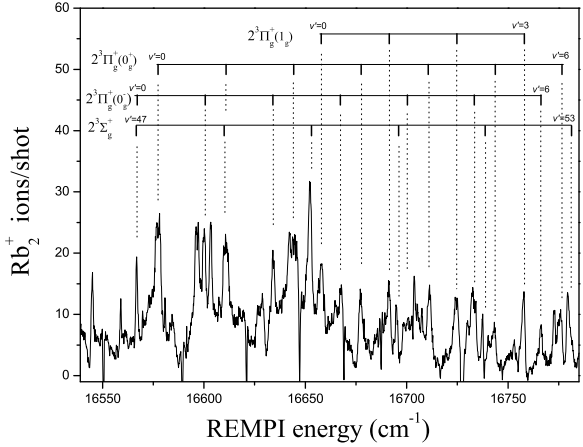}
\caption{\label{rempiv0spectrum}REMPI spectrum with the PA laser frequency tuned to the ($v'$=8, $J'=4$) level of the $1_g$ state. Tick marks above the spectrum are theoretical transition energies between the $a\,^3\Sigma_u^+$ $v''$=0 level and excited intermediate states $2\,^3\Sigma_g^+$ and $2\,^3\Pi_g(0_g^+, 0_g^-, 1_g)$. These transitions account for most of the observed lines.}
\end{figure}

To produce a REMPI spectrum, we set the PA laser frequency on a chosen rovibrational level and scan the REMPI laser. With the PA laser set to the $1_g$ ($v'$=8, $J'=4$) level, we obtain a REMPI spectrum (Fig. \ref{rempiv0spectrum}) which shows that $v''$=0 is present with a larger population than any other vibrational levels as predicted. Since the initial $a\,^3\Sigma_u^+$ state population is mostly in a single vibrational level, the REMPI spectrum is simplified and displays mostly the structure of the intermediate states ($2\,^3\Sigma_g^+$ and $2\,^3\Pi_g$) rather than a combination of initial and intermediate states.

The theoretical energies of REMPI transitions are calculated using the term energy of the $a\,^3\Sigma_u^+$ $v''$=0 level and the term energies of the intermediate states. The term energy of the $a\,^3\Sigma_u^+$ $v''$=0 level is calculated to be -234.73 cm$^{-1}$ using LEVEL 8.0 and the experimental  $a\,^3\Sigma_u^+$ potential \cite{strauss10}. The term energies of the intermediate states are calculated using LEVEL 8.0 and \textit{ab-initio} potentials \cite{park01} offset to match experimental data \cite{lozeille06}.

\section{Transition rates}
The PA laser creates molecules in the $1\,^3\Pi_g$ state, but these quickly decay to the $a\,^3\Sigma_u^+$ state where their lifetimes are orders of magnitude larger. The radiative lifetime of the $1\,^3\Pi_g$ state is $\sim 20$~ns. The lifetime of the $a\,^3\Sigma_u^+$ state, however, is not limited here by spontaneous emission but instead by the amount of time the molecules reside in the REMPI beam before before leaving ballistically  ($\tau_{transit}$ $\sim 5$~ms). Since there are at any given moment orders of magnitude more molecules in the $a\,^3\Sigma_u^+$ state than in the $1\,^3\Pi_g$ state, the former state dominates the ionization process. The measured number of Rb$_2^+$ ions is given by,

\begin{equation}
  N_{Rb_2^+}=   N_{a\,^3\Sigma_u^+} \cdot  p_{ionization} \cdot  e_d
\end{equation}
where $N_{Rb_2^+}$ is the number of $Rb_2^+$ ions measured per REMPI pulse, $N_{a\,^3\Sigma_u^+}$ is the steady state number of molecules in rovibrational levels of the $a\,^3\Sigma_u^+$ state that are resonant with the REMPI laser, $p_{ionization}$ is the photoionization probability per REMPI pulse, and $e_d$ is the efficiency of the ion detector.

Several rovibrational levels of the $a\,^3\Sigma_u^+$ state could be simultaneously resonant with different intermediate states whereby each would contribute to the total ion signal. However in the case of photoassociation to the $v'$=8 level of the $1_g$ state, only the $v''$=0 level contributes significantly to the ion signal.

Generally the first step of the REMPI process (bound-bound excitation) is fully saturated by the intense pulsed laser; however the second step of the REMPI process (bound-free photoionization) is not saturated and has a lower probability. We can calculate this ionization probability per pulse from,

\begin{equation}
  p_{ionization} =1-e^{W t }=1-e^{\frac{\sigma F}{t} t }=1-e^{\sigma E \lambda/(h c \; \pi w^2)}
\end{equation}
where the transition rate per second ($W=\frac{\sigma\,F}{t}$) is given by the photoionization cross section ($\sigma$) and the flux ($F$) per unit time. The flux ($F=E \lambda/(h c \; \pi w^2)$) is a measure of the total number of photons per unit area. $E$, $\lambda$, and $w$ are the pulse energy, wavelength and the Gaussian beam radius of the REMPI beam. $h$ and $c$ are Planck's constant and the speed of light. Although there are no known photoionization cross sections for the detection scheme we used, we can roughly estimate the cross section to be $\sigma$=$1_{-0.5}^{+5}$Mb based on other measurements \cite{Suemitsu83, creek68}.

Taking for example the transition to the $(v'$=8, J'=4) level of the $1_g$ state, we observe $N_{Rb_2^+}$=20 molecules per REMPI pulse. With the following set of parameters ($\sigma$=$1_{-0.5}^{+5}$~Mb, $E=5$~mJ, $\lambda=600$~nm, $w=1.4$~mm, $N_{Rb_2^+}$=20, and assuming $e_d=1$) we obtain $N_{a\,^3\Sigma_u^+}=90_{-60}^{+80}$ molecules residing within the REMPI beam volume in the steady state.

Solving the rate equation for the number of $a\,^3\Sigma_u^+$ state molecules, we obtain

\begin{equation}
   N_{a\,^3\Sigma_u^+}(t)=\frac{R_{PA} \cdot  FCF \cdot  t}{1+t/\tau}
\end{equation}
where $R_{PA}$ is the photoassociation rate per second, $\tau$ is the transit time lifetime of $a\,^3\Sigma_u^+$ molecules and $FCF$ is the fraction of $1\,^3\Pi_g$ molecules that decay to a particular vibrational level of the $a\,^3\Sigma_u^+$ state. For the typical parameter values of $FCF=0.37$ and $\tau_{transit}=5ms$, we obtain a photoassociation rate of $R_{PA}=5_{-1.5}^{+4}\times10^4$ molecules per second.

\section{Conclusions}
We have observed the formation of ultracold Rb$_2$ molecules by blue-detuned PA for the first time, as initially proposed in Ref.\cite{almazor01}. We have performed spectroscopy of the $1\,^3\Pi_g$ state and confirmed the double well structure of the $0_g^+$ state predicted in Ref. \cite{dulieu97}. We have shown that the $1_g$ $v'$=8 level mostly decays to the $a\,^3\Sigma_u^+$ $v''$=0 level. Extensions to the lowest rotational and hyperfine levels of the $a\,^3\Sigma_u^+$ state in Rb$_2$ and other alkali dimers may play an important role in cold chemistry and quantum information applications.

\section{Acknowledgements}
We gratefully acknowledge support from the NSF, AFOSR, and the UConn Research Foundation. A.G. acknowledges support from the Research Training Group 665 "Quantum interference and applications" Germany and from Laboratoire Aim\'e Cotton. Enlightening discussions with Johannes Deiglmayr and Fernand Spiegelman about the diabatization procedure are gratefully
acknowledged.


\footnotesize{
\bibliography{C:/Users/Public/bibliography/bibtex_masterfile_donotdelete}
\bibliographystyle{rsc} 
}

\end{document}